\documentclass[aps,twocolumn,epsfig,eqsecnum,showpacs]{revtex4}
\newcommand\be{\begin{eqnarray}}
\newcommand\ee{\end{eqnarray}}
\newcommand\ba{\begin{array}}
\newcommand\ea{\end{array}}

\def\r{\rangle}
\def\l{\langle}
\def\T{{\rm Tr}}

\def\cH{{\cal H}}
\def\cS{{\cal S}}
\def\cP{{\cal P}}

\def\cI{{\cal I}}
\def\cJ{{\cal J}}

\def\cF{{\cal F}}
\def\cU{{\cal U}}
\def\cE{{\cal E}}
\def\cO{{\cal O}}

\def\cV{{\cal V}}

\def\cM{{\cal M}}
\def\cN{{\cal N}}
\def\cA{{\cal A}}
\def\cB{{\cal B}}

\begin{document}
\title{Incomplete quantum process tomography and principle of maximal entropy}
\author{M\'ario Ziman}
\affiliation{Research Center for Quantum Information, Slovak Academy of Sciences, D\'ubravsk\'a cesta 9, 845 11 Bratislava, Slovakia}
\begin{abstract}
The main goal of this paper is to extend and apply
the principle of maximum entropy (MaxEnt) 
to incomplete quantum process estimation tasks. 
We will define a so-called
process entropy function being the von Neumann entropy of the 
state associated with the quantum process via Choi-Jamiolkowski isomorphism. 
It will be shown that an arbitrary process estimation experiment can be 
reformulated in a unified framework and MaxEnt principle can be consistently 
exploited. We will argue that the suggested choice for the process entropy
satisfies natural list of properties and it reduces to the 
state MaxEnt principle, if applied to preparator devices.
\end{abstract}

\maketitle


\section{Introduction}
Physical objects and processes are described by parameters that are directly,
or indirectly, accessible experimentally and represent the maximal knowledge
about physical systems (according to physical theory used). 
In quantum theory (see for instance \cite{peres,nielsen,zyckowski})
the complete information (knowledge) is represented by the concepts 
of quantum state (normalized positive operator), 
quantum observable (normalized positive operator valued measure) 
and quantum channel (completely positive linear trace-preserving map). 
One of the main characteristics of quantum system is its dimension $d$, i.e.
the maximal number of mutually perfectly distinguishable states
(in a single run of the experiment). These states form an orthogonal
basis of the associated complex Hilbert space $\cH$. 

An arbitrary quantum state is described as a positive trace-class 
linear operator with unit trace acting on the Hilbert space, 
$\varrho:\cH\to\cH:\varrho\ge 0,\T\varrho=1$, i.e. a density operator, 
or a density matrix. The number of independent real parameters 
determining the quantum states scales as $\cN_{\rm state}=d^2-1$. 
The quantum processes/operations correspond to completely positive 
trace-preserving linear maps defined on the set of all linear
operators including the set of all states $\cS(\cH)$. The number of 
independent real parameters determining the particular quantum operation 
equals $\cN_{\rm process}=d^2(d^2-1)$. Quantum measurements give us 
probability distributions over the set of all possible outcomes 
$\{x_1,\dots,x_L\}$, where $L$ is some positive integer. In theory,
the measured probabilities $p_j$ are determined by the Born's rule 
$p_j=\T \varrho F_j$, where $F_j$ is a positive operator (quantum effect) 
corresponding to outcome $x_j$. These operators form the so-called 
positive operator valued measure (POVM), i.e. $F_j$ are positive ($F_j\ge 0$) 
and they sum up to identity operator ($\sum_j F_j =I$).
The number of parameters specifying POVM depends on the total 
number of outcomes $L$ and equals $\cN_{\rm measurement}=(L-1)d^2$. 

The goal of quantum tomography is to estimate and fix all these
parameters \cite{rehacek_paris,nielsen_tomo,dariano}. 
However, already for small systems
(in dimension) the number of parameters is increasing rapidly, especially
for quantum channels \cite{altepeter,childs,obrien,weinstein,howard}. 
It seems that the complete knowledge about 
quantum objects is not a very realistic dream and experimentally 
we will not be able to perform all the desired tests 
\cite{gilchrist,emerson,lidar}. 
Fortunately, there
are situations in which even the knowledge of only few parameters
enables us to make reasonable and nontrivial predictions about the behavior
and properties of the system. A typical (classical) example
is the equilibrium thermodynamics in which only few parameters 
are used to describe the complex behavior of a system 
of approximately $10^{23}$ degrees of freedom. 
Our aim is to describe the properties of  
quantum objects as honestly as possible even in cases 
when the complete information is not available. In particular,
in this paper we will focus on incomplete quantum process tomography.

In Section II we will define the concept of process measurement
and shortly describe the idea of quantum process tomography. The maximum 
entropy principle is described in Section III and also the idea is
extended to process estimation problems by introducing the concept of
process entropy. Finally, in Section IV the MaxEnt procedure is 
applied to particular examples of incomplete ancilla-free estimation of 
qubit channels.

\section{Quantum process measurement}
A general quantum process tomography experiment consists of a test state
$\varrho$ that is transformed in some specific procedure $\cP$ involving
the unknown channel $\cE$ into a state $\varrho^\prime$, and a measurement
(POVM) $\cM$ performed on the state $\varrho^\prime$. This framework includes 
all the possible strategies \cite{nielsen_tomo,dariano,lidar} via 
which the parameters of quantum channels $\cE$ are accessible. We will 
define a {\it process measurement} as a particular choice of the test state 
$\varrho$, of the procedure $\cP$ and of the measurement $\cM$. Generally,
the procedure $\cP$ is composed of an application of some 
known quantum channels on the test state and one usage of the unknown 
channel $\cE$ acting on $d$-dimensional quantum system (qudit). That is 
$\cP$ itself is a quantum channel that can be written as
a product $\cP=\cP_{\rm in}\circ (\cF_{\rm anc}\otimes\cE)\circ\cP_{\rm out}$, 
where
$\cP_{\rm in},\cP_{\rm out}$ can be understood as being parts of the preparation
of the initial state $\varrho$, and of the final measurement $\cM$, 
respectively. Consequently, without loss of generality we can assume 
that $\cP=\cF_{\rm anc}\otimes\cE$, where $\cF_{\rm anc}$ is a 
known quantum channel acting on some ancillary system and also 
can be included as being a part of either preparation of $\varrho$, or 
measurement performed. A process measurement is called
ancilla-free if either the initial state $\varrho$ is factorized, 
or the ancillary system is trivial. Otherwise the process measurement
is ancilla-assisted and $\varrho^\prime=\cI_{\rm anc}\otimes\cE[\varrho]$.

For example, consider $\cE$ is a qudit quantum channel and the test state
$\varrho=\Psi_+$ is a maximally entangled state of two qudits 
($\Psi_+=\frac{1}{d}\sum |j\r\l k|\otimes|j\r\l k|$). 
The unknown channel is applied only on second of the qudits while 
the first one is transformed trivially, i.e. 
$\varrho^\prime=\omega_\cE=\cI\otimes\cE[\Psi_+]$ \cite{dariano}.
Performing the informationally complete POVM (resulting in complete
specification of $\omega_\cE$) the channel $\cE$ can be uniquely identified, 
because the mapping $\cE\mapsto\omega_\cE=\cJ[\cE]$ is the well known 
Choi-Jamiolkowski isomorphism \cite{choi,jamiolkowski} 
between the set of quantum qudit channels
and set of quantum states of two qudits. Hence, via general 
POVM measurements of the output state we can acquire 
either complete, or partial, knowledge on the channel. 
Let us note that individual ancilla-free process measurements 
cannot be informationally complete, but
they can be combined together to gain the complete information. In the
following sections we will concentrate onto situations 
in which the collection of process measurements provides us with partial
information, only.

\section{Principle of maximum entropy}
Maximum entropy (MaxEnt) principle was originally introduced in
statistics in order to estimate a probability distribution providing
that only partial information on that probability is available \cite{jaynes}.
There are many probability distributions compatible with the given
constraints and our aim is to choose one of them that in some sense 
represents our knowledge the most honestly. This choice cannot be logically
derived and some additional principle must be introduced.
Using the results of information theory \cite{shannon} 
on the uniqueness of Shannon entropy, one can argue that \cite{jaynes,jaynes61} 
the probability distribution maximizing the Shannon entropy
is the best choice we can make. Such probability maximizes the
uncertainty (measured by entropy) and, intuitively, also 
our predictions about the
unspecified parameters are as uncertain as possible. That is, a 
conclusion based on MaxEnt principle is introducing as little 
additional information as it is possible \cite{jaynes61}.

This idea was generalized to the domain of quantum state tomography 
\cite{blankenbecler,derka}
by using the concept of von Neumann entropy \cite{peres} 
$S(\varrho)=-\T\varrho\log\varrho$, which is considered to be the 
quantum extension of Shannon entropy. A state observation level $\cO_n$
is defined as a set of $n$ ($n\le d^2$) mean values $\{f_1,\dots,f_n\}$
of linearly independent operators $\{F_1,\dots,F_n\}$ 
related to unknown state $\varrho$ via the trace rule 
$f_j=\T\varrho F_j=\l F\r_\varrho$.
If the operators $\{F_1,\dots,F_n\}$ form a POVM the numbers $f_j$ represent
the measured probabilities. The answer to incomplete state tomography 
problem based on maximum entropy principle is given by the following equation
\be
\nonumber
\varrho=\arg \max \{S(\varrho)| \varrho\in\cS(\cH), 
f_j = \T\varrho F_j, j=1,\dots, n\}\, .
\ee
The following state is the formal solution of the MaxEnt problem 
\cite{jaynes,derka}
\be
\varrho=\frac{1}{Z}\exp(-\sum_j \lambda_j F_j)\, ,
\ee
where $Z=\T[\exp(-\sum_j \lambda_j F_j)]$ and $\lambda_j$ are 
Lagrange multipliers fixed by the system of equations 
\be
f_j=\T\varrho F_j=-\frac{\partial}{\partial \lambda_j} 
\ln Z(\lambda_1,\dots,\lambda_n)\, .
\ee

For example, consider a two-level quantum system (qubit)
and observation levels 
\be
\nonumber
& & \cO_1=\{\l\sigma_z\r_\varrho\}\\ \nonumber
& &\quad\subset\cO_2=\{\l\sigma_y\r_\varrho,\l\sigma_z\r_\varrho\} \\ \nonumber
& & \quad\quad\subset \cO_3=\{\l\sigma_x\r_\varrho,\l\sigma_y\r_\varrho,\l\sigma_z\r_\varrho\}\, .
\ee
A qubit state can be expressed in a so-called Bloch sphere picture 
as  $\varrho=\frac{1}{2}(I+\vec{r}\cdot\vec{\sigma})$ with
$r_j=\T\varrho\sigma_j$. The MaxEnt principle applied for $\cO_1,\cO_2$ 
sets mean values $r_j$ of all the unobserved operators to zero.
That is, for $\cO_1=\{z\}$ we get $\varrho=\frac{1}{2}(I+z\sigma_z)$
and for $\cO_2=\{y,z\}$ the MaxEnt estimation gives
$\varrho=\frac{1}{2}(I+y\sigma_y+z\sigma_z)$. Observation level $\cO_3$
provides complete information about the quantum state, hence the principle
of maximum entropy is not needed in this case.

Clearly, there is a problem if we consider similar incomplete
estimation task for processes, namely, which entropy should
be maximized? Unlike quantum states the quantum channels are lacking 
some concept of entropy, or uncertainty. In fact, what does it mean
that a quantum process is uncertain? Our goal is to introduce a suitable
concept of a channel/process entropy $S_{\rm proc}(\cE)$ 
and investigate its properties. Before analyzing
different choices let us discuss some (intuitive) properties
of the process entropy. 
\begin{enumerate}
\item{\it Uncertainty of unitary channels.}
Without any doubts the unitary channels play a very specific role
among all quantum processes. For unitary processes the interaction
of the system with its environment is trivial. 
The physical invertibility is the unique and characteristic
property of the unitary channels. In some sense the channel entropy
should reflect how much noise the channel introduces. 
Unitary channels are noiseless and in what follows we will assume 
that the channel entropy is invariant 
under unitary preprocessing ($\cV$) and unitary postprocessing ($\cU$), i.e., 
$S_{\rm proc}(\cE)=S_{\rm proc}(\cU\circ\cE\circ\cV)$. It follows that
all unitary channels have the same value of uncertainty that can be 
set to zero. Moreover, we do require that $S_{\rm proc}(\cE)=0$
implies that $\cE$ is unitary.

\item{\it Uniqueness of maximum.} For the purposes of 
incomplete process estimation exploiting the MaxEnt principle 
it is necessary that the maximum is unique. Hence there must be 
a unique channel, for which the uncertainty is maximal. This channel
should be the result of the incomplete estimation if no data are 
available, i.e., when the observation level is trivial, $\cO_0=\emptyset$.
Because of the unitary invariance the channel must be invariant under
unitary preprocessing and postprocesing, i.e., 
$\cE_{\max}=\cU\circ\cE_{\max}\circ\cV$. Only the channel mapping
the whole state space into a total mixture ($\varrho\mapsto\frac{1}{d}I$) 
is invariant in this sense. It is argued in \cite{ziman2005} that this channel 
is indeed the average channel over all possible qubit channels. To guarantee
that in any process measurement the maximum is unique it is sufficient
that the process entropy is a concave function, i.e., 
$S_{\rm proc}(\lambda\cE_1+[1-\lambda]\cE_2)\ge\lambda S_{\rm proc}(\cE_1)
+[1-\lambda] S_{\rm proc}(\cE_2)$.

\item{\it Universality.}
This is not a condition on the concept of process entropy itself, but rather
on the general possibility to employ such principle once we agree on a suitable
measure of channel entropy. It is important that the maximum entropy principle 
is applicable for all process measurements. We shall discuss this issue later 
in more details.

\end{enumerate}

In summary, a channel entropy is some concave function 
(defined on the set of quantum channels) achieving its maximum
for the complete contraction to the total mixture and vanishing
only for unitary channels. A natural choice of process entropy seems to be
related to the concepts of quantum channel capacity 
\cite{schumacher, westmoreland, holevo, kretschmann}. 
Quantum capacity quantifies the degree of preservation of quantum 
states during the transmission and this value is different for different
unitary transformations. On the other hand, the classical capacity
is maximal also for noisy channels. For example, phase-damping channels 
$\varrho\to{\rm diag}[\varrho]$ maximize the transmission of classical
information over the quantum channels. Because of these properties, the
capacities are not appropriate candidates for the definition of process 
entropy usable in incomplete process tomography tasks. 

\subsection{Choi-Jamiolkowski process entropy}
The Choi-Jamiolkowski isomorphism provides us naturally with a notion
of channel entropy. It uniquely associates a quantum state 
$\omega_\cE=(\cI\otimes\cE)[\Psi_+]$ with a quantum channel $\cE$, 
hence we can adopt the
von Neumann entropy of $\omega_\cE$ as being the channel entropy of $\cE$
\cite{zyczkowski2004,zyczkowski2008}.
Consider a quantum channel on $d$-dimensional system (qudit).
Providing that for each process measurement we are able to define uniquely
a state observation level $\cO_n$ given by mean values $x_j=\T X_j\omega_\cE$ 
of $n$ linearly independent Hermitian operators $X_j$, 
the MaxEnt problem for processes can be formalized as follows
\be
\nonumber
\cE=\arg \max_{\omega_\cE}S(\omega_\cE)
\ee
where the maximum of von Neumann entropy $S(\omega_\cE)$ is taken over all
states $\omega_\cE\in\cS(\cH\otimes\cH)$ satisfying the constraints
$\T_2\omega_\cE=\frac{1}{d}I$ and $x_j = \T\omega_\cE X_j$ for all 
$X_1,\dots,X_n\in\cO_n$. The resulting state $\omega_\cE$ 
determines the quantum operation $\cE$ uniquely via the inverse
relation
\be
\cE[\varrho]=d\T_{\rm anc}[(\varrho^T\otimes I)\omega_\cE]\, .
\ee 
In what follows we will
investigate the process entropy given as the von Neumann entropy of
the state $\omega_\cE$ associated with the channel $\cE$ via 
Choi-Jamiolkowski formalism.

It is straightforward to see that only for unitary channels the states
$\omega_\cE$ are pure and hence $S_{\rm proc}(\cE)=S(\omega_\cE)=0$
only for unitary channels, $\cE=\cU$. Moreover, unitary channels 
do not change the entropy of $\omega_\cE$, i.e., 
$S(\omega_\cE)=S(\omega_{\cU\circ\cE\circ\cV})$. The concavity of $S_{\rm proc}$
follows from the concavity of von Neumann entropy and the maximum is achieved
for $\omega_\cE=\frac{1}{d^2}I$ that is associated with the channel 
mapping the whole state space into the maximally mixed state,
$\cE:\varrho\mapsto\frac{1}{d}I$. In summary, the process entropy 
\be
\label{eq:process_entropy}
S_{\rm proc}(\cE)=-\T \omega_\cE\log\omega_\cE
\ee
satisfies all the desired properties we have discussed previously. The only
open issue is its applicability in general process measurement. 

Choi-Jamiolkowski isomorphism is associated with a specific
process measurement using as the test state a maximally entangled 
state of two qudits. Second qudit is sent through the unknown channel 
while the first one is evolving trivially to obtain the state
$\cI\otimes\cE[\Psi_+]=\omega_\cE$, that is estimated in some state 
measurement described by POVM. In this case the process observation 
level can be defined as the following set of mean values
\be
\cO^{\rm proc}_n=\{x_1,\dots,x_n\}\, ,
\ee
where
\be
x_j=\T F_j\omega_\cE=\l F_j\r_{\cI\otimes\cE[\Psi_+]}\equiv \l F_j\r_\cE\; .
\ee
Because of the identity $\T_2\omega_\cE=\frac{1}{d}I$
the process observation level is equivalent to a state observation level
\be
\nonumber
\cO_{n+d^2-1}=\{x_1,\dots,x_n,0,\dots,0\}\, .
\ee
The added zeros represent the mean values
of $d^2-1$ operators $I\otimes\Lambda_j$, where
$\Lambda_j$ are traceless Hermitian qudit operators
forming a basis of the set of traceless Hermitian qudit operators, i.e.,
the general qudit state can be written as 
$\varrho=\frac{1}{d}(I+\vec{r}\cdot\vec{\Lambda})$. 
To be more precise we assume that the operators $F_1,\dots, F_n$ 
are linearly independent of operators $I\otimes\Lambda_1,\dots,
I\otimes\Lambda_{d^2-1}$.

What if the test state is not the maximally entangled one?
Is it possible to interpret the measured values as linear constraints
on the state $\Omega_\cE$ defined by Choi-Jamiolkowski isomorphism?
Let us note that the linearity is crucial, because we implicitly
assume that the constraints representing the incomplete information
are linear, which guarantees that the set of possible solutions
is convex and, hence, the entropy has a unique maximum.

\subsection{General quantum process experiment {\it vs} Choi-Jamiolkowski 
isomorphism}
Consider a general test state $\Omega$ of the qudit and an
arbitrary ancilla system. We will show that there exist a 
completely positive linear map $\cA_\Omega:\cB(\cH_d)\to\cB(\cH_{\rm anc})$ 
such that $\cA_\Omega\otimes\cI[\Psi_+]=\Omega$. A general pure state
$|\Phi\r=\sum_{\alpha, j} \Phi_{\alpha j}|\alpha\r_{\rm anc}\otimes|j\r$
($\alpha=1,\dots, D$; $j=1,\dots d$)
can be written as $|\Phi\r= A_{\Phi}\otimes I|\Psi_+\r$, where 
the operator $A_\Phi:\cH\to\cH_{\rm anc}$ is defined as
$A_\Phi=\sqrt{d}\sum_{\alpha, j} \Phi_{\alpha j} |\alpha\r\l j|$.
A general mixed state $\Omega$ can be written as convex combination
of pure states $\Omega=\sum_k \lambda_k |\Phi_k\r\l \Phi_k|$, hence
$\Omega = \sum_k \lambda_k (A_j\otimes I)\Psi_+(A_j^\dagger\otimes I)
=(\cA_\Omega\otimes\cI)[\Psi_+]$. Since the values $\lambda_k$ are positive
the transformation $\cA_{\Omega}$ is completely positive. Moreover,
for each state $\Omega$ the linear map $\cA_\Omega$ is unique.
Hence, for a general test state $\Omega$ the mean value of
an Hermitian operator $F$ can be expressed as follows
\be
\nonumber
\l F\r_{(\cI\otimes\cE)[\Omega]}&=&\l F\r_{(\cA_\Omega\otimes\cE)[\Psi_+]}\\
&=&
\l (\cA^*_\Omega\otimes\cI)[F] \r_{(\cI\otimes\cE)[\Psi_+]}\, , 
\label{eq:reduction}
\ee
where $\cA^*_{\Omega}$ is a dual mapping to $\cA_{\Omega}$ (Heisenberg picture).
As a result we get that an arbitrary ancilla-assisted process measurement 
can be rewritten within the framework of process 
measurement using the maximally entangled
test state $\Psi_+$ and measuring a suitable 
Hermitian operator $\cA^*_\Omega\otimes\cI[F]$, hence, the maximum
entropy principle defined via Choi-Jamiolkowski entropy
can be consistently employed in all incomplete process measurements.

In what follows we shall analyze the ancilla-free process
measurement, hence only the qudit itself is used to probe the action
of the quantum channel. In fact, this can be considered
as an ancilla-assisted problem with a factorized test state
$\Omega=\xi_{\rm anc}\otimes\varrho$, and a factorized measurement 
resulting in the mean value of the operator of the form $I_{\rm anc}\otimes F$. 
Consider $\Omega$ is a pure factorized
state $|\Phi\r=|\varphi\r_{\rm anc}\otimes|\psi\r$. Then the operator $A_\Phi$
takes the following form $A_\Phi=|\varphi\r\l \psi^*|$, where $|\psi^*\r$
is a complex conjugated state, i.e., $\l k|\psi^*\r = \overline{\l k|\psi\r}$
for all basis vectors $|k\r$, in which the maximally entangled state $\Psi_+$
is defined. It follows that for general factorized state 
$\Omega=\sum_{\alpha,l}\lambda_\alpha \mu_l |\varphi_\alpha\r\varphi_\alpha|
\otimes |\psi_l\r\l\psi_l |$ the transformation $\cA_\Omega$ is
expressed via Kraus operators $A_{\alpha l}=\sqrt{d\lambda_\alpha \mu_l} 
|\varphi_\alpha\r\l\psi_l^*|$. Therefore, according to Eq.~(\ref{eq:reduction})
the ancilla-free process measurement of $I_{\rm anc}\otimes F$
can be considered as an ancilla process measurement with the maximally entangled
state $\Psi_+$ being the test state and a measurement of
$X=(\cA^*_\Omega\otimes\cI)[I_{\rm anc}\otimes F]$, i.e.,
\be
\nonumber
X&=&
\sum_{\alpha,l} 
(A^\dagger_{\alpha l}\otimes I)(I_{\rm anc}\otimes F)(A_{\alpha, l}\otimes I)\\
\nonumber &=&
d\sum_{\alpha,l} \lambda_\alpha \mu_l 
|\psi^*_l\r\l\varphi_\alpha|\varphi_\alpha\r\l\psi^*_l|\otimes F\\
\nonumber &=&
d (\sum_l \mu_l |\psi^*_l\r\l\psi^*_l|)\otimes F\\
\nonumber &=&
d\varrho^T\otimes F\, ,
\ee
where we used that $\varrho=\sum_l \mu_l |\psi_l\r\l\psi_l| $,
$\xi_{\rm anc}=\sum_\alpha \lambda_\alpha |\varphi_\alpha\r\l\varphi_\alpha|$
and $\varrho^T$ is the transposed matrix $\varrho$ with respect to
basis $\{|k\r\}$. 

We have shown that measuring the outcome
associated with $F$ in the ancilla-free process measurement
is equivalent to measuring $d\varrho^T\otimes F$ in
the process measurement with maximally entangled state $\Psi_+$, where
$\varrho$ is the ancilla-free test state. It means that 
the ancilla-free process observation level consisting of mean
values $\l F_1\r_{\varrho_1}\dots,\l F_n\r_{\varrho_n}$
is equivalent to $\cO_n^{\rm proc}=\{\l d\varrho_1^T\otimes F_1\r_\cE,
\dots,\l d\varrho_n^T\otimes F_n\r_\cE\}$.

\subsection{States as preparation channels}
Preparation devices play a completely different role than quantum channels. 
However, formally, they can be understood as mappings that transform an 
arbitrary input state into a fixed output state $\xi$. In this sense
{\it preparation channels} $\cE_\xi$
form a very specific convex subset of quantum channels. Let us apply
the proposed maximum entropy based process tomography to preparation
channels, i.e., to preparators. The process measurement is ancilla-free
consisting of all linearly independent test states 
$\varrho_j$ ($j=1,\dots,d^2$).
and measurement of the mean value of Hermitian operator $F$. 
According to previous paragraph the process observation level is described as 
$\cO_{d^2}^{\rm proc}=\{\l d\varrho_1^T\otimes F\r_{\cE_\xi},
\dots,\l d\varrho_{d^2}^T\otimes F\r_{\cE_\xi}\}$.
The Choi-Jamiolkowski entropy of the channel $\cE_\xi$
equals (up to a constant) to the von Neumann entropy of the state $\xi$,
because $\cI\otimes\cE_\xi[\Psi_+]=\frac{1}{d}I\otimes\xi$ implies
$S(\omega_{\cE_\xi})=\log_2 d+S(\xi)$. Moreover, because of the identity
\be
\l d\varrho_j^T\otimes F\r_{\cI\otimes\cE_\xi[\Psi_+]}
=\l d\varrho_j^T\otimes F\r_{\frac{1}{d}I\otimes\xi}=\l F\r_{\xi}\,,
\ee
it follows that finding a channel with the maximal Choi-Jamiolkowski entropy  
is equivalent to finding a state maximizing the von Neumann entropy.
As a result we get that the process MaxEnt procedure, if applied to 
preparators, reduces to the state MaxEnt procedure. That is,
the proposed Choi-Jamiolkowski process entropy is a consistent 
extension of the von Neumann entropy. In particular, the MaxEnt principle 
for states can be considered as being a special case
of the MaxEnt principle for channels.

\section{Examples}
In this section we shall present few examples of incomplete quantum process
estimation for ancilla-free process measurements of a qubit channel. 
\subsection{$\cO_1^{\rm proc}=\{\l 2\varrho^T\otimes\sigma_z\r_\cE\}$}
In this case the collected data provides us about information
on the mean value of an observable $\sigma_z$, hence the experiment
gives us single value $m=\l\sigma_z\r_{\cE[\varrho]}$. Unfortunately, even 
in this simplest case we cannot give (see Appendix A) an analytic 
solution in its whole generality. In particular, we found the solutions
in following cases
\be
\begin{array}{lcl}
\varrho=\frac{1}{2}I &:& \cE[\xi]=\frac{1}{2}(I+m\sigma_z)\, ,\\
\varrho=|\psi\r\l\psi| &:& \cE[\xi]=\frac{1}{2}(I+\frac{1}{2}m(1+(\vec{t}\cdot\vec{r})\sigma_z)\, ,
\end{array}
\ee
where $\varrho=\frac{1}{2}(I+\vec{r}\cdot\vec{\sigma})$
and $\xi=\frac{1}{2}(I+\vec{t}\cdot\vec{\sigma})$. It is interesting that
for pure test state the estimated channel is not unital even if there
are unital channels satisfying the constraints. 

An alternative method for incomplete process estimation was described in
\cite{ziman2005}. It is based on a different {\it ad hoc} rule demanding 
that no additional information about unobserved measurements (those completing
the incomplete process observational level) is introduced.  In particular, 
for states $\eta$ orthogonal (in Hilbert-Schmidt sense, 
i.e., $\T \eta\varrho=0$) to given test states, the mean values are 
completely random, i.e., they are transformed into the total mixture 
($\eta\to\frac{1}{2}I$). Hence the entropy of output states for
unmeasured inputs is maximal. It means that if possible 
(meaning there is no contradiction with the data, or theory) the 
total mixture is preserved. Otherwise an optimization procedure minimizing 
the average distance from the total mixture is needed. This method was
analyzed only for qubit channels and for ancilla-free process measurements.
The extension of the method to all process measurements
will require introduction of additional rules. Let us compare
the method proposed in \cite{ziman2005} and the one proposed in this paper.

If measuring $\sigma_z$ and finding $m=\pm 1$ the output state 
must be pure and it corresponds to an eigenvalue of $\sigma_z$. 
In both mentioned scenarios we know the solution providing 
our knowledge consists of complete information of the action 
of the channel on the pure test state, thus, we know that
$\cE:|\psi\r\mapsto|\pm z\r$, respectively.
As it was argued in the work \cite{ziman2005} the estimated transformation
should map the whole Bloch sphere into the line connecting north 
and south pole, i.e., $\vec{t}\to\vec{t}^\prime=(0,0,\pm t_z)$. 
However, the proposed MaxEnt estimation procedure 
gives different result. In particular, 
$\vec{t}\to \vec{t}^\prime=(0,0,\pm (1+t_z)/2)$. This transformation
is not unital, but the total mixture is mapped to the state 
$\vec{t}^\prime=(0,0,1/2)$. A state $\vec{t}=-\vec{r}$ 
orthogonal to the test state $\vec{r}$ is transformed as follows
\be
\label{eq:example}
\begin{array}{lll}
\mathrm{ Scheme\ in\ [22]}: & m=1\quad 
& \cE_{\rm est}:-\vec{r}\mapsto -\vec{r} \\
{\rm MaxEnt:} & m=1\quad  & \cE_{\rm est}:-\vec{r}\mapsto\vec{0}
\end{array}
\ee
As we see in this case both methods transform orthogonal 
(in Hilbert-Schmidt sense) states to $|\psi\r$ into the total mixture, 
but for MaxEnt procedure also the orthogonal (in Hilbert space sense) 
state $|\psi_\perp\r$ is mapped into the total mixture. 
In our opinion this feature (except the universality) 
justifies the usage of MaxEnt procedure 
in comparison with the scheme described in 
\cite{ziman2005}. In fact, the uncertainty introduced by the estimation 
procedure on perfectly distinguishable (orthogonal) states from
the test states should be as maximal as possible. And this is not the case
for the method used in \cite{ziman2005}, for which the estimated channel 
preserves the orthogonal state.

\subsection{$\cO_3^{\rm proc}=
\{\l I\otimes\sigma_x\r_\cE,
\l I\otimes\sigma_y\r_\cE,\l I\otimes\sigma_z\r_\cE\}$}
Consider a situation that the unknown qubit channel is tested by the total
mixture and the complete tomography of the output state is performed,
i.e., mean values of $\sigma_x,\sigma_y,\sigma_z$ are known. 
The corresponding state observation level is 
$\cO_6=\{\l\vec{\sigma}\otimes I\r_\varrho,
\l I\otimes\vec{\sigma}\r_{\varrho}\}=\{\vec{0},\vec{m}\}$,
for which the solution is presented in Appendix B. In such case the proposed
MaxEnt process tomography procedure leads us to the channel
\be
\cE_{\rm est}:\varrho\mapsto\varrho_0=\frac{1}{2}(I+\vec{m}\cdot\vec{\sigma})\, ,
\ee
hence, the whole Bloch sphere is contracted into a single point 
$\vec{m}$. As a result we get that if the total mixture is used to probe 
the channel action then according to MaxEnt procedure all the states are
mapped into the output state $\varrho_0=\cE[\frac{1}{2}I]$. In this case both
the discussed procedures are giving the same estimation.

\subsection{$\cO_4^{\rm proc}=
\{\l I\otimes\sigma_z\r_\cE,\l 2(|x\r\l x|)^T\otimes\sigma_z\r_\cE,
\l 2(|y\r\l y|)^T\otimes\sigma_z\r_\cE,\l 2(|z\r\l z|)^T\otimes \sigma_z\r_\cE\}$}
In this case the process is probed with four test states
(total mixture and positive eigenvectors of $\sigma_x,\sigma_y,\sigma_z$
forming a vector of pure states $\vec{\eta}$),
but only $z$th component of the Bloch vector of the output state
is measured. $\cO_7=\{\l I\otimes\sigma_z\r_\varrho, 
\l 2\vec{\eta}^T\otimes\sigma_z\r_\varrho,\l\vec{\sigma}\otimes I\r_\varrho\}=
\{z,\vec{\zeta},\vec{0}\}$ is the corresponding state estimation
problem and $z,\vec{\zeta}$ are the experimentally identified 
mean values. Information encoded in these parameters can be equivalently 
rewritten into the form $\cO_7=\{\l I\otimes\sigma_z\r_\varrho, 
\l\vec{\sigma}\otimes\sigma_z\r_\varrho, \l\vec{\sigma}\otimes I\r_\varrho\}
=\{z,\vec{\zeta}^\prime,\vec{0}\}$, where $\zeta_j^\prime=\zeta_j -z$.

It is shown in \cite{derka} that for such state observation level
the estimated density matrix reads
\be
\omega=\frac{1}{4}\left(
I\otimes I
+z I\otimes\sigma_z
+(\vec{\zeta}^\prime\cdot\vec{\sigma})\otimes\sigma_z
\right)\, .
\ee
Hence, the process is described by the following state transformation
($\varrho\to\cE_{\rm est}[\varrho]$)
\be
\vec{t}\to\vec{t}^\prime=(0,0,z+\vec{\zeta}^\prime\cdot\vec{t})\, ,
\ee
i.e., $\varrho^\prime=\frac{1}{2}
[I+(z+\vec{\zeta}^\prime\cdot\vec{t})\sigma_z]$.
As in all previous cases, also in this case the whole state space is mapped
onto a subset of the line connecting states $|+z\r$ and $|-z\r$. However, 
in this case the final state depends also on parameters $t_x,t_y$.

\section{Conclusion and discussion}
We have addressed the problem of incomplete process estimation
based on maximum entropy principle \cite{jaynes61}.
In general the maximum entropy principle is an intuition-based 
{\it ad hoc} principle related to quantification of 
ignorance contained in probability distributions that seems to agree
with our experience. This ignorance measured in entropy can be extended 
to domain of quantum states by introducing the von Neumann entropy. Our 
attempt here was to develop similar approach for processes. We argued that
capacities are not good candidates for quantifying the uncertainty of
quantum channels and we exploited the Choi-Jamiolkowski process entropy
defined as
\be
\label{ch_entropy}
S_{\rm proc}(\cE)=-\T [\omega_\cE\log\omega_\cE]\, , \quad
\omega_\cE=(\cI\otimes\cE)[\Psi_+]\, ,
\ee
where $\Psi_+$ is the maximally entangled state. In particular,
we showed that the suggested concept can be universally applied in all
possible process measurements. The procedure is demonstrated 
on three incomplete ancilla-free estimation problems of a qubit channel: i) 
pure test state and projective measurement, ii) the total mixture
as the test state and complete tomography of the output state, and iii)
four test states and the same projective measurement.

We have shown that unlike the concepts of capacity of quantum channels
the process entropy defined above is compatible with the following
properties:
\begin{enumerate}
\item{\it Uniqueness of maximum:} unique maximum for the channel contracting 
whole state space into the total mixture.
\item{\it Minimum:} minimum is achieved only for unitary processes.
\item{\it Unitary invariance:} invariant under unitary transformations, i.e.,
$S_{\rm proc}(\cU\circ\cE\circ\cV)=S_{\rm proc}(\cE)$ for all unitary 
transformations $\cU,\cV$.
\item{\it Concavity:} function $S_{\rm proc}(\cE)$ is concave, i.e.,
$S_{\rm proc}(p\cE+q\cF)\ge qS_{\rm proc}(\cE)+pS_{\rm proc}(\cF)$.
\end{enumerate}
The proposed Choi-Jamiolkowski process entropy serves as a very valuable
tool in incomplete process tomography deserving future testing and
investigation. Moreover, as it is shown in Section III.B, the 
proposed process entropy principle, if applied to state preparator 
devices, is equivalent to the state entropy principle based 
on von Neumann entropy. The key feature discussed in this manuscript
is the universality of the proposed procedure following from
the unification of all process measurements described in Section III.B.
This idea goes beyond the applications in incomplete process estimation
and is further developed in \cite{ziman_ppovm}.

Recently, Olivares et al. \cite{olivares} proposed and analyzed 
a state estimation problem combining incomplete information with
some nontrivial apriori knowledge. In their approach the maximization 
of entropy is replaced by minimization of Kullback relative entropy
$S(\varrho|\varrho_0)=\T[\varrho(\log\varrho-\log\varrho_0)]$
with a bias $\varrho_0$ representing the prior knowledge. This approach
can be directly extended to the case of channels by introducing
the quantity
$S(\omega_\cE||\omega_0)=-\T[\omega_\cE(\log\omega_\cE-\log\omega_0)]$
with $\cE_0$ playing the role of prior information. If we set $\cE_0$
to be the state space contraction into the total mixture
(i.e. $\omega_0=\frac{1}{d^2}I$), then
$\T[\omega_\cE\log\omega_0]=-\log{d^2}\T\omega_\cE=-2\log d$.
Consequently, $S(\omega_\cE||\omega_0)=2\log d-S(\omega_\cE)$ and the 
biased estimation problem reduces to the unbiased maximum process 
entropy estimation.

Let us give a simple example based on the observation level
discussed in Section IV.A. Suppose that out of the performed measurement 
we acquire the information $|0\r\mapsto|0\r$. We shall consider three 
different priors: 
{\it i)} identity channel $\cE_0=\cI$;
{\it ii)} diagonalisation channel $\cE_0={\rm diag}$ transforming each
state into its diagonal form in the basis $|0\r,|1\r$;
{\it iii)} 
$\cE_0=\frac{1}{2}(\cI+\cU_x)$, where $\cU_x[\xi]=\sigma_x\xi\sigma_x$.
Let us note that $S(\varrho||\varrho_0)$ is finite only if
the support of $\varrho$ is included in the support of $\varrho_0$.
For the case of identity channel $\omega_0$ is a pure state, hence
$S(\omega_\cE||\omega_0)<\infty$ only if $\omega_\cE=\omega_0$, i.e.
$\cE_{\rm est}=\cI$. Fortunately, the identity channel is in accordance 
with the constraint $|0\r\mapsto|0\r$, hence the estimation gives the 
identity channel. In the second case it is straightforward to verify 
that the channel ${\rm diag}$ fullfils the constraints. Since 
$S({\rm diag}||\cE_0)=0$ is the minimal possible value we get 
$\cE_{\rm est}={\rm diag}$. In the third case the support of
$\omega_0$ is a linear span of vectors 
$|\psi_+\r=(|00\r+|11\r)/\sqrt{2}$ and
$|\phi_+\r=(|01\r+|10\r)/\sqrt{2}$. Therefore, only for channels
with $\omega_\cE = a|\phi_+\r\l\phi_+|+b|\psi_+\r\l\psi_+|$ the relative entropy is finite. However, the constraint requires
that $|0\r\l 0|=(a\cI+b\cU_x)[|0\r\l 0|]=a|0\r\l 0|+b|1\r\l 1|$, i.e.
necessarily $b=0$. In such case $a=1$, because otherwise $a\cI$ is not
a valid quantum channel. That is, only the identity channel satisfies the
measured constraint, thus it minimizes the Kullback relative entropy. 
The estimation gives $\cE_{\rm est}=\cI$. In all these cases we find 
different estimations as in the unbiased maximum entropy approach 
(see Eq.(\ref{eq:example})).
The role of prior information in incomplete process estimation
deserves much deeper analysis than it is presented in these simple 
examples. However, such task is beyond the scope of this manuscript.

\acknowledgements This work was supported by in part by the European
Union  projects QAP,  by the Slovak Academy of Sciences 
via the project CE-PI and by the Slovak grant agency APVV and VEGA.
\appendix
\section{MaxEnt solution for $\cO_1^{\rm proc}=
\{\l2\varrho^T\otimes\sigma_z\r_\cE\}$ 
for qubit channels}
According to Section III this process observation level is equivalent
to the following state observation level
\be
\nonumber
\cO_4&=&\{\l 2\varrho^T\otimes\sigma_z\r_\varrho,\l\sigma_x\otimes I\r_\varrho, \l\sigma_y\otimes I\r_\varrho, \l\sigma_z\otimes I\r_\varrho\}
\\ \nonumber &=&\{m,0,0,0\} \, .
\ee
Maximum entropy estimation gives us the following state
\be
\omega=\frac{1}{Z}\exp{[-\vec{\lambda}\cdot(\vec{\sigma}\otimes I)-2d\varrho^T\otimes\sigma_z]}
\ee
where $Z=\T[\exp[-\vec{\lambda}\cdot(\vec{\sigma}\otimes I)-2d\varrho^T\otimes\sigma_z]]$ and $\vec{\lambda}, d$ are Lagrange multipliers 
that can be determined by solving the system of algebraic equations
\be
\label{o1_lagrange}
\vec{0}=-\frac{\partial}{\partial\vec{\lambda}}\ln Z \ \ \ \ 
m=-\frac{\partial}{\partial d}\ln Z \, .
\ee
Using the expression $\varrho^T=\frac{1}{2}(I+\vec{r}_T\cdot\vec{\sigma})$
($\vec{r}_T=(r_x,-r_y,r_z)$) the state can be written in the form 
$\omega=\frac{1}{Z}e^{-R}$ with
\be 
R&=& A\otimes|0\r\l 0|+B\otimes|1\r\l 1|\\
\nonumber
A&=&[(\vec{\lambda}+d\vec{r}_T)\cdot\vec{\sigma}+dI]\\
\nonumber
B&=&[(\vec{\lambda}-d\vec{r}_T)\cdot\vec{\sigma}-dI]\, .
\ee
Since the operators $A\otimes|0\r\l 0|$ and $B\otimes|1\r\l 1|$
commute we can write
\be
e^{-R} &=& e^{-A\otimes |0\r\l 0|}e^{-B\otimes |1\r\l 1|}\\
\nonumber
&=& e^{-A} \otimes |0\r\l 0|+e^{-B} \otimes |1\r\l 1|\, .
\ee
Having in mind the operator identity
\be
e^{xI+\vec{y}\cdot\vec{\sigma}}=
e^x(\cosh|\vec{y}|+\frac{\sinh|\vec{y}|}{|\vec{y}|}\vec{y}\cdot\vec{\sigma})\, 
\ee
we obtain
\be
\nonumber
Z=\T e^{-R}=2(e^{-d} \cosh|\vec{\lambda}+d\vec{r}_T|
+e^{d} \cosh|\vec{\lambda}-d\vec{r}_T|)\, .
\ee
Inserting this expression into Eqs.(\ref{o1_lagrange}) we get
\be
\nonumber
\vec{0}&=& S_+\frac{(\vec{\lambda}+d\vec{r}_T)}{|\vec{\lambda}+d\vec{r}_T|}+
S_-\frac{(\vec{\lambda}-d\vec{r}_T)}{|\vec{\lambda}-d\vec{r}_T|}\\
m (C_+ + C_-)
&=& C_+ -C_- -2S_+ \frac{(\vec{\lambda}+d\vec{r}_T)\cdot\vec{r}_T}
{|\vec{\lambda}+d\vec{r}_T|} 
\label{ex1_conditions}
\ee
where $S_\pm = e^{\mp d}\sinh|\vec{\lambda}\pm d\vec{r}_T|$
and $C_\pm = e^{\mp d}\cosh|\vec{\lambda}\pm d\vec{r}_T|$.
From the first of these equations it follows that $\vec{\lambda}+d\vec{r}_T$
and $\vec{\lambda}-d\vec{r}_T$ are collinear, i.e.,
$\vec{\lambda}+d\vec{r}_T=k(\vec{\lambda}-d\vec{r}_T)$. This is possible only
if either $\vec{\lambda}=0$, or $d=0$, or $\vec{\lambda}=\lambda\vec{r}_T$. 

The case $d=0$ requires that $\vec{\lambda}=\vec{0},m=0$
whatever test state $\vec{r}_T$ is used. Thus, measuring
the mean value $\l\sigma_x\r=m=0$ the MaxEnt results in the
state $\omega = \frac{1}{4}I\otimes I$ and, consequently,
the estimated channel acts as follows
\be
\xi\to\xi^\prime = 2\T_1[(\xi^T\otimes I)\omega]=\frac{1}{2}I\, ,
\ee
i.e., the whole Bloch sphere is transformed into the total mixture.

For the case $\vec{\lambda}=\vec{0}$ the first equation implies
that either $d=0$ (leads to same solution as before), or 
$\vec{r}_T=\vec{0}$. If the the test state is chosen to be in total
mixture ($\vec{r}_T=\vec{0}$), the second equation
leads to $d=-{\rm arctanh}(m)$, hence the estimated state reads 
\be
\nonumber
\omega&=&\frac{1}{4\cosh d}e^{-d I\otimes\sigma_z}
\\ &=& \frac{1}{4\cosh d}I\otimes(\cosh d-\sinh d \sigma_z)
\nonumber
\\ &=& \frac{1}{2}I\otimes\frac{1}{2}(I+m\sigma_z)\, .
\ee
The corresponding process $\cE_{\rm est}$ is given by the identity
$\cE[\xi]=d\T_1 [(\xi^T\otimes I)\omega]$, i.e.,
\be
\cE_{\rm est}[\xi]&=&\frac{1}{2}\T_1 
[(\xi^T\otimes I)(I\otimes (I+m\sigma_z))]
\\ &=& \frac{1}{2}\T_1 [\varrho^T\otimes(I+m\sigma_z)]
\\ &=& \frac{1}{2}(I+m\sigma_z) \, .
\ee
This transformation maps the whole state space
into the single point $\xi=\frac{1}{2}(I+m\sigma_z)$.

The last family of solutions of MaxEnt conditions 
is that the vectors $\vec{\lambda}$ and $\vec{r}_T$ are collinear. 
In this case we reduced the number of unknown parameters 
to $\lambda=|\vec{\lambda}|$ and $d$. The first condition out of 
Eqs.~(\ref{ex1_conditions}) then reads
\be
\nonumber
0&=& e^{-d}\frac{\sinh|(\lambda+d)r|}{|\lambda+d|}
(\lambda+d)+
e^d \frac{\sinh|(\lambda-d)r|}{|\lambda-d|}
(\lambda-d)
\ee
where we used $r=|\vec{r}|=|\vec{r}_T|$. Analyzing all possible
values for $\lambda\pm d$ it follows that the absolute values 
can be omitted and the equations simplify to
\be
\label{eq_1}
0&=& e^{-d}s_+ +e^d s_-\\
\label{eq_2}
m&=& \frac{e^{-d}c_+ -e^dc_- -2re^{-d}s_+}
{e^{-d}c_+ +e^d c_-}\, ,
\ee
where $s_\pm = \sinh[(\lambda\pm d)r]$, $c_\pm = \cosh[(\lambda\pm d)r]$.
After a short algebra these equations can be rewritten into the form
\be
e^{\lambda r}\cosh[d(1-r)]=e^{-\lambda r}\cosh[d(1+r)]
\ee
and
\be
\nonumber
& & m(e^{\lambda r}\cosh[d(1-r)]+e^{-\lambda r}\cosh[d(1+r)])=\\
\nonumber
& & \ \ = -e^{\lambda r}\sinh[d(1-r)]-e^{-\lambda r}\sinh[d(1+r)])\\
& & \ \ \ \ \ -2re^{-d}\sinh[(\lambda+d)r]\, .
\ee

Unfortunately, we cannot give a general solution in a closed form. 
Consider therefore a special case and let us assume that the test state 
is pure, i.e., $r=1$. In such case the solution reads
\be
d=\frac{1}{2}{\rm arctanh}(-m)=\frac{1}{4}\ln\frac{1-m}{1+m}
\ee
\be
\lambda = \frac{1}{2}\ln\cosh(2d) \, .
\ee
As a result we get
\be
\nonumber
\omega&=&
\frac{1}{Z}e^{-d}[\cosh(\lambda+d)I
-\sinh(\lambda+d)\vec{r}_T\cdot\vec{\sigma}]\otimes|0\r\l 0|\\
\nonumber & & +\frac{1}{Z}e^d[\cosh(\lambda-d)I
-\sinh(\lambda-d)\vec{r}_T\cdot\vec{\sigma}]\otimes|1\r\l 1|
\ee
with $Z=2(e^\lambda+e^{-\lambda}\cosh(2d))$.
Let us denote by $\vec{t}$ the Bloch vector corresponding to a
general input state $\xi$, then the estimated operation is given by 
the following prescription
\be
\nonumber
\xi\to\xi^\prime &=& 2\T_1 [(\xi^T\otimes I)\omega]\\
\nonumber
&=& \frac{1}{2}(1+\frac{1}{2}m(1+\vec{t}_T\cdot\vec{r}_T))|0\r\l 0|\\
& & +\frac{1}{2}(1-\frac{1}{2}m(1+\vec{t}_T\cdot\vec{r}_T))|1\r\l 1|\,
\ee
where we have used $\xi=\frac{1}{2}(1+\vec{t}\cdot\vec{\sigma})$.
Taking into account that $\vec{t}_T\cdot\vec{r}_T=\vec{t}\cdot\vec{r}$
we can write
\be
\xi\to\xi^\prime=
\frac{1}{2}(I+\frac{1}{2}m(1+\vec{t}\cdot\vec{r})\sigma_z)\, .
\ee
In the language of Bloch vectors the transformation reads 
\be
\vec{t}\to \vec{t}^\prime=(0,0,\frac{1}{2}m[1+\vec{t}\cdot\vec{r}])\, .
\ee

\section{MaxEnt solution for 
$\cO_6=\{\l I\otimes\vec{\sigma}\r_\varrho,\l\vec{\sigma}\otimes I\r_\varrho\}
=\{\vec{m},\vec{0}\}$} 
According to maximum entropy principle the state maximizing the entropy
has the form 
$\omega=\frac{1}{Z}e^{-(\vec{\lambda}\cdot\vec{\sigma})\otimes I-
I\otimes(\vec{\mu}\cdot\vec{\sigma})}$ with
\be
Z&=&\T e^{-(\vec{\lambda}\cdot\vec{\sigma})\otimes I-
I\otimes(\vec{\mu}\cdot\vec{\sigma})}\\
\nonumber &=&
(\T e^{-\vec{\lambda}\cdot\vec{\sigma}})(\T e^{-\vec{\mu}\cdot\vec{\sigma}})
=4\cosh|\vec{\lambda}|\cosh|\vec{\mu}|\, .
\ee
The values of $\vec{\lambda},\vec{\mu}$ are given by 
the following system of equations
\be
\vec{0}=-\frac{1}{Z}\frac{\partial Z}{\partial\vec{\lambda}}
&\Rightarrow&
\vec{0}=-\tanh|\vec{\lambda}|\frac{\vec{\lambda}}{|\vec{\lambda}|} \\
\vec{m}=-\frac{1}{Z}\frac{\partial Z}{\partial\vec{\mu}}
&\Rightarrow&
\vec{m}=-\tanh|\vec{\mu}|\frac{\vec{\mu}}{|\vec{\mu}|}\, ,
\ee
and for the estimated state we get
\be
\omega=\frac{1}{Z}e^{-(\vec{\lambda}\cdot\vec{\sigma})\otimes I-
I\otimes(\vec{\mu}\cdot\vec{\sigma})}
=\frac{1}{4}I\otimes(I+\vec{m}\cdot\vec{\sigma})\, .
\ee
As a result we found that the estimated channel $\cE_{\rm est}$ 
maps the whole Bloch sphere into the point 
$\frac{1}{2}(I+\vec{m}\cdot\vec{\sigma})$.



\begin{thebibliography}{00}

\bibitem{peres}
  A.Perez: {\it Quantum Theory: Concepts and Methods},
  (Kluwer, Dordrecht, 1993)
\bibitem{nielsen}
  M.A. Nielsen and I.L. Chuang:
  {\it Quantum Computation and Quantum Information},
  (Cambridge University Press, Cambridge, 2000)
\bibitem{zyckowski}
  I.Bengtsson and K.Zyczkowski:
  {\it Geometry of quantum states: An introduction to quantum entanglement},
  (Cambridge University Press, Cambridge, 2006)
\bibitem{rehacek_paris}
  {\it Quantum State Estimation}, edited by M.Paris and J.\v Reh\'a\v cek,
  (Springer, 2004)
\bibitem{nielsen_tomo}
  I.L.Chuang, and M.A.Nielsen
  J.Mod.Phys. 44, 2455-2467 (1997), [arXive: quant-ph/9610001]
\bibitem{dariano}
  G.M.D'Ariano and P.L.Presti,
  Phys. Rev. Lett. 91 047902-1 (2003), [arXive:quant-ph/0211133]
\bibitem{altepeter}
  J.B.Altepeter, {\it et al.},
  Phys.Rev.Lett. 90, 193601 (2003)
\bibitem{childs}
  A.M.Childs, I.L.Chuang, and D.W.Leung,
  Phys.Rev.A 64, 012314 (2001)
\bibitem{obrien}
  J.O'Brien, {\it et al.},
  Phys.Rev.Lett. 93, 080502 (2004)
\bibitem{weinstein}
  Y.S.Weinstein, {\it et al.},
  J.Chem.Phys. 121, 6117-6133 (2004)
\bibitem{howard}
  M.Howard, {\it et al.},
  New J.Phys. 8, 33 (2006)
\bibitem{gilchrist}
  A.Gilchrist, N.K.Langford, and M.A.Nielsen,
  Phys.Rev.A 71, 062310 (2005)
\bibitem{emerson}
  J.Emerson, R.Alicki, K.Zyczkowski,
  J.Opt.B: Quantum Semiclass. Opt. 7, S347 (2005)
\bibitem{lidar}
  M.Mohseni, and D.A.Lidar,
  Phys. Rev. Lett. 97, 170501 (2006) [arXive:quant-ph/0601033 ]
\bibitem{choi}
  M.Choi,
  Linear Algebra and Its Applications, 285-290 (1975)
\bibitem{jamiolkowski}
  A.Jamiolkowski, 
  Rep. Math. Phys. 3, 275 (1972)
\bibitem{jaynes}
  E.T.Jaynes:
  {\it  Probability Theory: The Logic of Science},
  (Cambridge University Press, Cambridge, 2003)
\bibitem{shannon}
  C.E.Shannon, 
  Bell System Tech. J. 27, 379-423, 623-656 (1948)
\bibitem{jaynes61}
  E.T.Jaynes, 
  in Statistical Physics, K. Ford (ed.), Benjamin, New York, p. 181 (1963)
\bibitem{blankenbecler}
  R.Blankenbecler, and M.H.Partovi,
  Phys.Rev.Lett. 54, 373 (1985)
\bibitem{derka}
  V.Bu\v zek, G.Drobn\'y, G.Adam, R.Derka, and P.L.Knight,
  J. Mod. Optics 44, 2607 (1997)
\bibitem{ziman2005}
  M.Ziman, M.Plesch, and V.Bu\v zek :
  Foundations of Physics 36, 127-156 (2006), [arXive:quant-ph/0406088]
\bibitem{schumacher}
  B.Schumacher, M.A.Nielsen, 
  Phys.Rev.A 54, 2629 (1996)
\bibitem{westmoreland}
  B.Schumacher, M.D.Westmoreland,
  Phys.Rev.A 56, 131 (1997)
\bibitem{holevo}
  A.S.Holevo,
  IEEE Trans.Inf.Theory 44, 269 (1998), [quant-ph/9611023]
\bibitem{kretschmann}
  D.Kretschmann, R.Werner, 
  New J. Phys. 6,  26 (2004), [arXive:quant-ph/0311037]
\bibitem{zyczkowski2004}
  Karol Zyczkowski, Ingemar Bengtsson,
  Open Sys. and Information Dyn. 11, 3-42 (2004)
\bibitem{zyczkowski2008}
  Wojciech Roga, Mark Fannes, and Karol Zyczkowski, 
  J.Phys.A 41, 035305 (2008)
\bibitem{olivares}
  S.Olivares, M.G.A.Paris,
  Phys.Rev.A 76, 042120 (2008)
\bibitem{ziman_ppovm}
  M.Ziman, 
  Phys.Rev.A 77, 062112 (2008), [arXiv:0802.3862]
\end{thebibliography}
\end{document}